\input{aipcheck}

\documentclass[
    ,final            
  ]
  {aipproc}

\layoutstyle{6x9}

\begin{document}

\title{A 350-MHz GBT Survey of 50 Faint {\it Fermi} $\gamma$-ray Sources for Radio Millisecond Pulsars}


\author{J.~W.~T. Hessels}{
  address={ASTRON, Postbus 2, 7990 AA Dwingeloo, The Netherlands}
  ,altaddress={Astronomical Inst., Univ. of Amsterdam, 1098 SJ Amsterdam, The Netherlands}
}

\author{M.~S.~E. Roberts}{
  address={Eureka Scientific, Inc., Oakland, California 94602, USA}
}

\author{M.~A. McLaughlin}{
  address={Department of Physics, West Virginia University, 210 Hodges Hall, Morgantown, WV 26506, USA}
  ,altaddress={Also adjunct at the National Radio Astronomy Observatory, Green Bank, WV 24944, USA}
}

\author{P.~S. Ray}{
  address={Space Science Division, Naval Research Laboratory, Washington, DC 20375-5352, USA}
}

\author{P. Bangale}{
  address={Department of Physics, West Virginia University, 210 Hodges Hall, Morgantown, WV 26506, USA}
}

\author{S.~M. Ransom}{
  address={NRAO, 520 Edgemont Road, Charlottesville, Virginia 22093, USA}
}

\author{M. Kerr}{
  address={Department of Physics, University of Washington, Seattle, WA 98195-1560, USA}
}

\author{F. Camilo}{
  address={Columbia Astrophysics Laboratory, Columbia University, New York, NY 10027, USA}
}

\author{M.~E. DeCesar}{
  address={Department of Astronomy, University of Maryland, College Park, MD 20742, USA}
}

\author{the Fermi PSC}{
  address={Fermi Pulsar Search Consortium}
}

\begin{abstract}
We have used the Green Bank Telescope at 350\,MHz to search 50 faint, unidentified {\it Fermi} $\gamma$-ray sources for radio pulsations.  So far, these searches have resulted in the discovery of 10 millisecond pulsars, which are plausible counterparts to these unidentified {\it Fermi} sources.  Here we briefly describe this survey and the characteristics of the newly discovered MSPs.
\end{abstract}

\maketitle


\section{Introduction}

Since the launch of the {\it Fermi Gamma-Ray Space Telescope}, it has become clear that a large fraction of Galactic $\gamma$-ray sources are not only pulsars, but nearby millisecond pulsars (MSPs).  This has come somewhat, though not completely, as a surprise and has spurred a coordinated effort to deeply search the positions of {\it Fermi} sources for radio MSPs.  So far this combined effort has discovered over 30 radio MSPs\footnote{These searches are coordinated under the ``{\it Fermi} Pulsar Search Consortium".}, which are also likely to emit pulsed $\gamma$-ray emission.  These new MSPs are predominantly very nearby ($d \sim 1$\,kpc), making multi-wavelength optical and X-ray follow-up feasible.  Radio timing observations are underway and will serve to construct precise rotational ephemerides in order to detect $\gamma$-ray pulsations from many of these sources \citep[e.g.,][]{rrc+10}.

Here we present the discovery of 10 of these MSPs, which were found in a 350-MHz Green Bank Telescope (GBT) survey of 50 faint, unidentified {\it Fermi} sources\footnote{These are sources {\it not} included in the {\it Fermi} ``Bright Source List", which encompassed sources identified at 10-sigma or greater confidence during the first 3 months of operation.} .  We briefly describe the survey setup and the characteristics of the newly discovered MSPs.

\begin{figure}[t] 
   \centering
   \includegraphics[width=6in]{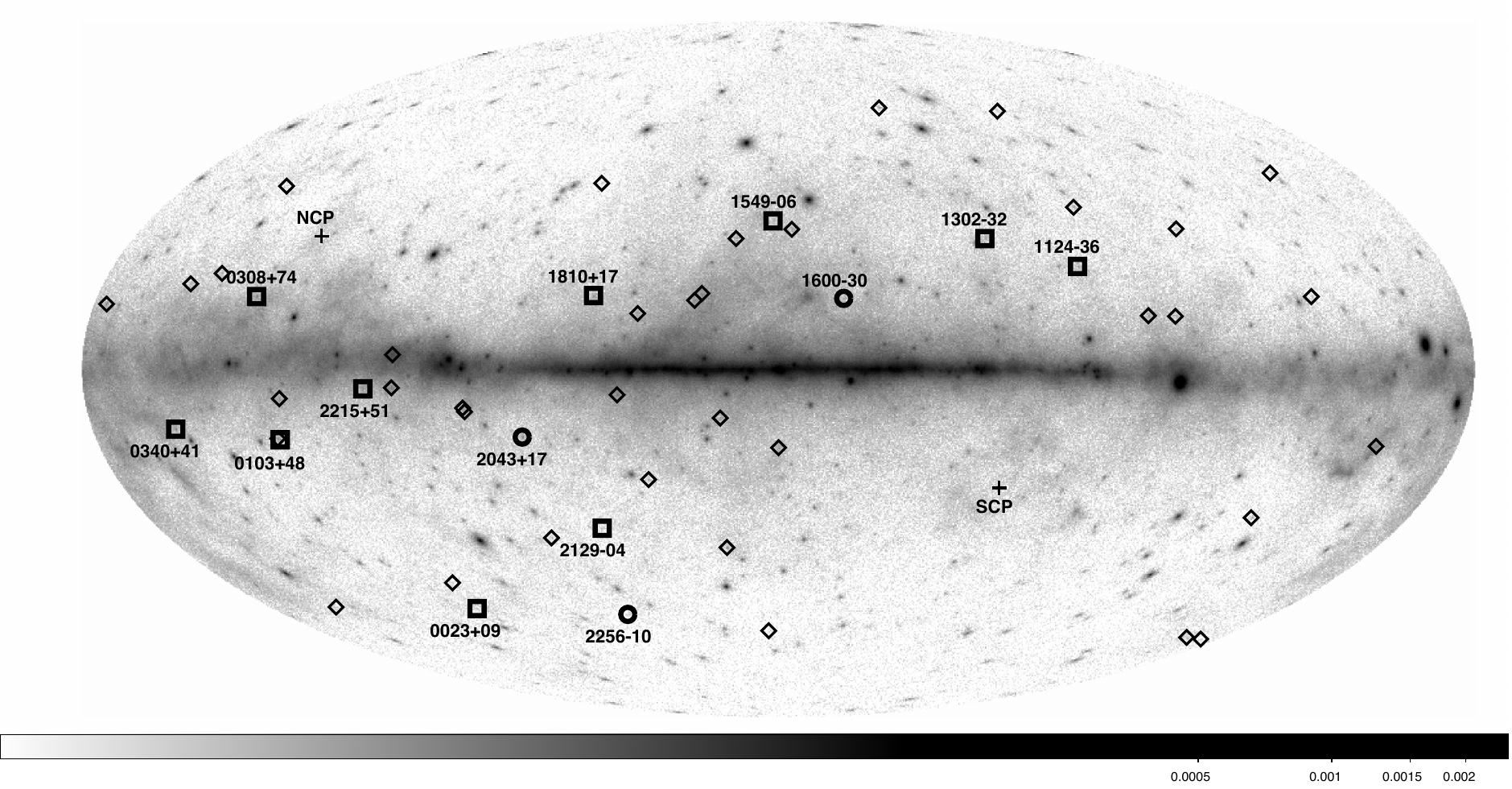} 
   \caption{The positions of the 50 sources surveyed here, overlaid on a {\it Fermi} map of the $\gamma$-ray sky in Galactic coordinates.  The north (NCP) and south (SCP) celestial poles are marked for reference.  Sources with (as yet) no new pulsar detections are marked with diamonds (note that one source position is partially obscured by the marker of the nearby source 0103+48).  Those with previously known or newly discovered MSPs are marked with circles and boxes respectively and are labelled with the pulsar name.}
   \label{fig:allskymap}
\end{figure}

\section{Observations and Analysis}

We used the GBT in combination with the Green Bank Ultimate Pulsar Processor (GUPPI) backend at a central frequency of 350\,MHz.  GUPPI recorded a 100-MHz bandwidth, spanned by 4096 spectral channels and recorded with a time resolution of 81.92\,$\mu$s.  This system is extremely sensitive to nearby pulsars - those with low DM and scattering measure, and especially those with steep spectral indices\footnote{For example, one of the new pulsar discoveries, J0308+74, was too faint to see at 1.4\,GHz in a 1-hr observation with the 100-m Effelsberg telescope.} - and is also conducive to blind periodicity searches because of the generally low level of radio frequency interference (RFI) contaminating the band.  The beam full-width half max is 0.6\,deg, sufficiently large to encompass the positional uncertainty of the faintest {\it Fermi} sources, which can be on the order of 0.1-0.3\,deg.  

In October/November 2009, we observed 50 {\it Fermi} catalog sources (Figure~\ref{fig:allskymap}), typically with an integration time of 32 minutes.
We specifically targeted sources away from the Galactic plane ($|b| > 5$\,deg), where sky temperature and scattering are reduced and generally pose little problem even at 350\,MHz.  In most cases, these targeted observations provided a factor of 5 increase in sensitivity over past wide-field 350-MHz surveys of this part of the sky.  The observed $\gamma$-ray sources had no obvious blazar identification and we preferentially observed sources that were identified as being ``pulsar-like" on the basis of their MeV$-$GeV spectra and low flux variability.  

The data were analyzed with standard Fourier-based acceleration search techniques (to search for pulsars in compact binaries), investigating DM trials ranging from 0\,pc\,cm$^{-3}$ up to twice the NE2001 model predicted total Galactic DM in the direction of each source \citep{cl02}.  Candidates were sifted to identify/merge harmonically related signals and to excise interference.  The highest-ranked candidates were then folded to produce diagnostic plots for further consideration as potential pulsars.

\begin{figure}[t] 
   \centering
   \includegraphics[width=6in]{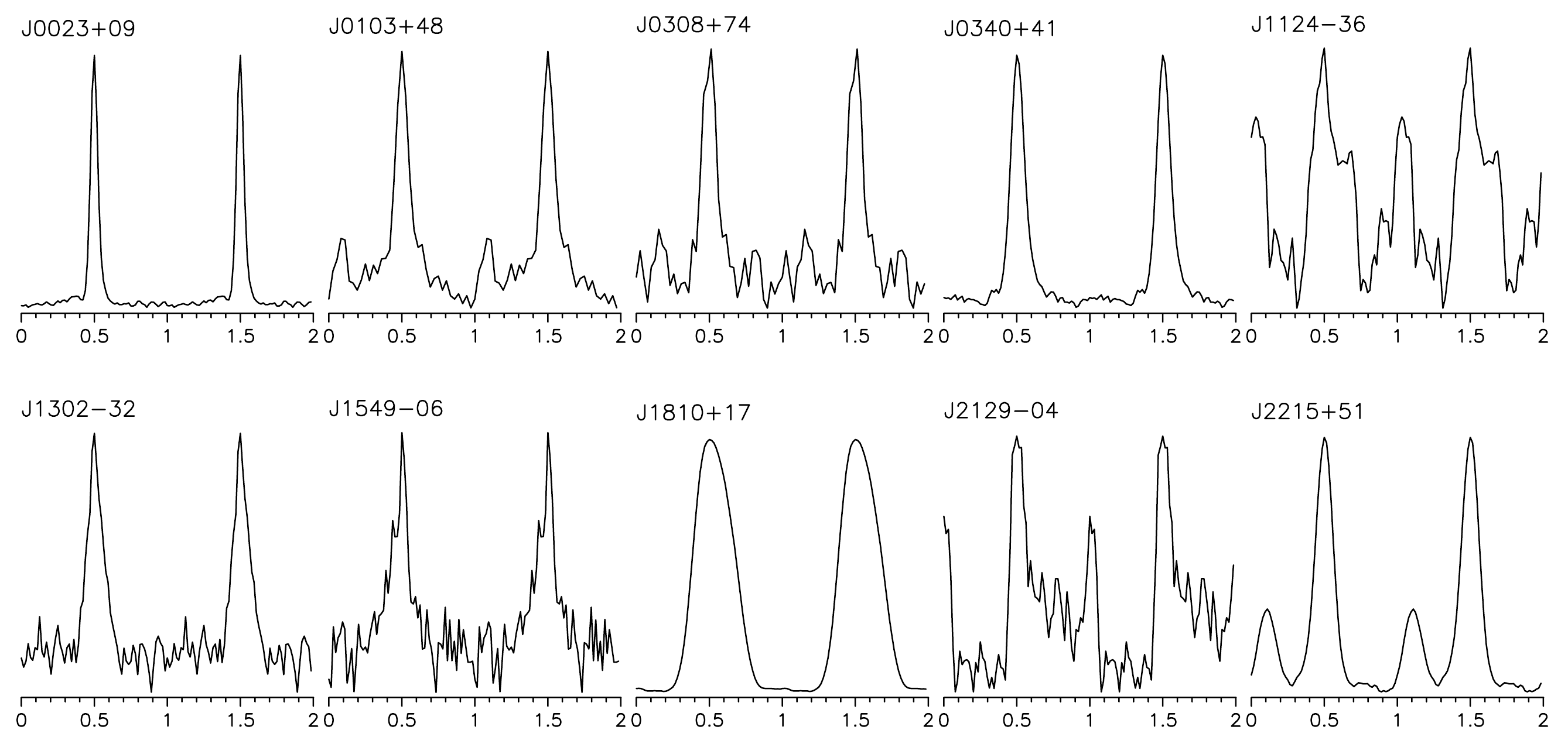} 
   \caption{The cumulative pulse profiles of the 10 new pulsars.  For clarity, the pulse profiles are repeated over two complete rotations.}
   \label{fig:profiles}
\end{figure}

\section{Results}

With the data from $> 80$\% of the 50 sources thoroughly searched\footnote{We have also searched the first 200\,s of all data sets in order to quickly identify the brightest sources and/or those found in particularly compact orbits ($P_{\rm orb} \sim 0.5$\,hr).}, we have found 10 new MSPs and confirmed 1 recent MSP discovery within the survey region (J2043+17, found first by the Nan\c{c}ay Telescope in a similar targeted survey of {\it Fermi} sources).  The pulse profiles of these new sources are shown in Figure~\ref{fig:profiles}.  The currently known characteristics of these pulsars are summarized in Table~\ref{tab:pulsars}, which also includes two previously known MSPs coincident with survey sources but found in earlier, untargeted surveys (J1600$-$3053 and J2256$-$1023).  Given the very recent discovery of some sources, not all source parameters are yet known.  It is worth noting that the majority of these newly discovered pulsars are too faint to have been found in recent GBT 350-MHz surveys of the northern sky, which have limiting sensitivities of roughly $S_{400} \sim 2$\,mJy \citep[e.g.,][]{hes08}.  Including the previously known MSPs coincident with our survey targets, the success rate for finding radio MSPs coincident with this particular set of $\gamma$-ray sources currently stands at 13/50, an impressive success rate.

At least 7 of the sources are in a binary system.  At least 3 of these pulsars, J0023+09, J1810+17, and J2215+51, are in compact, few-hour orbits in which the pulsar is eclipsed for part of the orbital period.  J0023+09 and J1810+17 appear to be classical ``black-widow" pulsars, with very low-mass, presumably degenerate companions.  In contrast, J2215+51 has a possibly non-degenerate $\sim 0.2$\,M$_{\rm Sun}$ companion and appears similar in its orbital characteristics to the recently discovered X-ray binary / MSP ``missing link", PSR~J1023+0038 \citep{asr+09}.  These eclipsing systems are discussed in more detail in \citep{rob11}.

All of these pulsars are being followed-up in order to derive phase-connected rotational ephemerides which can be used to fold the {\it Fermi} photon data and to determine whether these sources also pulse in $\gamma$-rays.  The isolated pulsar J0340+41 is the first to have a complete timing solution and $\gamma$-ray pulsations have been detected \citep{ban11}.  Similar analysis will soon be possible for several of the other sources, though a few of the new discoveries are quite weak and will require significant observing time in order to derive a full phase-connected timing solution.  Several of the binary sources are being followed-up at optical and X-ray wavelengths.

\begin{table}[t]
\caption{New MSPs Found in a 350-MHz GBT Survey of {\it Fermi} $\gamma$-ray Sources}
\begin{tabular}{llcccccccl}
\hline
\tablehead{1}{l}{b}{Pulsar\tablenote{Pulsars in italics were previously known.  PSR J2256$-$1023 was found in the GBT 350-MHz drift-scan survey and will be presented by \citep{sta11}.}} & \tablehead{1}{l}{b}{1FGL} & \tablehead{1}{c}{b}{P$_{\rm spin}$} & \tablehead{1}{c}{b}{DM} & \tablehead{1}{c}{b}{P$_{\rm orb}$} & \tablehead{1}{c}{b}{M$^{\rm min}_{\rm c}$} & \tablehead{1}{c}{b}{$S_{350}$} & \tablehead{1}{c}{b}{DM Dist.\tablenote{Estimated using the NE2001 model of \citep{cl02}.}} \\
            &            &  (ms)                   & (pc cm$^{-3}$) & (hr)                   & (M$_{\rm Sun}$)           & (mJy)          & (kpc)       \\
\hline
J0023+09     & J0023.5+0930     & 3.05 & 14.3 & 3.3          & 0.016   & 2 & 0.7 \\
J0103+48     & J0103.1+4840     & 2.96 & 53.5 & 40             & 0.18        & 0.5 & 2.3 \\
J0308+74     & J0308.6+7442     & 3.16 &   6.35 & Binary?  & ?        & 0.3 & 0.6 \\
J0340+41     & J0340.4+4130     & 3.29 & 49.6 & Isolated   & N/A        & 2 & 1.8 \\
J1124$-$36  & J1124.4$-$3654 & 2.41 & 44.9 & Binary      & ?        & 0.3 & 1.7 \\
J1302$-$32  & J1302.3$-$3255 & 3.77 & 26.2 & $> 24$   & ?        & 0.5 & 1.0 \\
J1549$-$06 & J1549.7$-$0659  & 7.09 & 21.6 & Binary?    & ?        & 0.5 & 1.0 \\
{\it J1600$-$3053} & {\it J1600.7$-$3055}  & 3.60  &  52.3 & 343  & 0.20 & ?  & 1.6 \\
J1810+17     & J1810.3+1741     & 1.66 & 39.7 & 3.6          & 0.045   & 20 & 2.0 \\
J2043+17\tablenote{Discovered first with the Nan\c{c}ay Telescope.} & J2043.2+1709     & 2.38 & 20.7 & Binary      & ?             & 0.8 & 1.8 \\
J2129$-$04 & J2129.8$-$0427  & 7.62  & 16.9 & Binary     & ?              & 0.5 & 0.9 \\
J2215+51     & J2216.1+5139     & 2.61 & 69.2 & 4.2         & 0.22      & 5 & 3.0 \\
{\it J2256$-$1023} & {\it J2256.9$-$1024}  & 2.29 & 13.8 & 5 & 0.03 & 7 & 0.6 \\
\hline
\end{tabular}
\label{tab:pulsars}
\end{table}%


\begin{theacknowledgments}
J.W.T.H. is a Veni Fellow of The Netherlands Organisation for Scientific Research (NWO).  This research was partially funded through the Fermi GI program, NASA grant \#NNG10PB13P.
\end{theacknowledgments}



\bibliographystyle{aipproc}   

\bibliography{Hessels_Sardinia2010_Proc_v3}

\IfFileExists{\jobname.bbl}{}
 {\typeout{}
  \typeout{******************************************}
  \typeout{** Please run "bibtex \jobname" to optain}
  \typeout{** the bibliography and then re-run LaTeX}
  \typeout{** twice to fix the references!}
  \typeout{******************************************}
  \typeout{}
 }

\end{document}